\documentclass[preprint, 10p]{elsarticle}

\usepackage{microtype}
\usepackage{graphicx}
\usepackage{color}
\usepackage{url}
\usepackage{multirow}
\usepackage{booktabs}
\usepackage{amssymb,amsmath,amsthm,amsfonts}
\usepackage[caption=false]{subfig}

\usepackage{lineno,hyperref}
\modulolinenumbers[5]

\usepackage{xcolor, soul}

\newcommand{\return}{\vspace{0.2cm}}

\DeclareMathOperator{\polylg}{polylog}
\newcommand{\RNum}[1]{\uppercase\expandafter{\romannumeral #1\relax}}

\newcommand{\var}[1]{\emph{#1}}

\newtheorem{prop}{Property}
\newtheorem{thm}{Theorem}
\newtheorem{lem}{Lemma}
\newtheorem{cor}{Corollary}
\newtheorem{problem}{Problem}

\newtheorem{obs}{Observation}

\newcommand{\seq}{\mathcal{S}}
\newcommand{\ef}{\textup{\textsf{EF}}}
\newcommand{\bound}{\textup{\textsf{B}}}

\newcommand{\access}{\textup{\textsf{Access}}}
\newcommand{\ins}{\textup{\textsf{Insert}}}
\newcommand{\del}{\textup{\textsf{Delete}}}
\newcommand{\pred}{\textup{\textsf{Predecessor}}}
\newcommand{\suc}{\textup{\textsf{Successor}}}

\newcommand{\append}{\textup{\textsf{Append}}}

\newcommand{\upd}{\textup{\textsf{Update}}}
\newcommand{\sm}{\textup{\textsf{Sum}}}
\newcommand{\search}{\textup{\textsf{Search}}}
\newcommand{\minimum}{\textup{\textsf{Minimum}}}
\newcommand{\maximum}{\textup{\textsf{Maximum}}}
\newcommand{\select}{\textup{\textsf{Select}}}
\newcommand{\rank}{\textup{\textsf{Rank}}}

\newcommand{\addr}{\textup{\textsf{Address}}}
\newcommand{\realloc}{\textup{\textsf{Realloc}}}

\newcommand{\bit}[1]{\texttt{#1}}
\newcommand{\gray}[1]{{\color{gray}{#1}}}

\makeatletter
\def\@subsecfont{\bfseries}
\renewcommand\subsection{\@startsection{subsection}{3}{0pt}%
  {-.5\baselineskip \@plus -2\p@ \@minus -.2\p@}%
  {3.5\p@}%
  {\@subsecfont}}
\makeatother

\journal{}

\begin{document}

%
%
%

\begin{frontmatter}

\title{Succinct Dynamic Ordered Sets with Random Access\tnoteref{conference}}

\author{Giulio Ermanno Pibiri\fnref{address1}}
\author{Rossano Venturini\fnref{address2}}

\fntext[address1]{ISTI-CNR, Italy. Email address: giulio.ermanno.pibiri@isti.cnr.it}

\fntext[address2]{University of Pisa and ISTI-CNR, Italy. Email address: rossano.venturini@unipi.it}

\tnotetext[conference]{A preliminary version of this work was published in the Proceedings of the 28-th Annual Symposium on Combinatorial Pattern Matching (CPM 2017)~\cite{PV17}, when the first author was a Ph.D. student
at the University of Pisa, Italy. \\
This work was partially supported by the BIGDATAGRAPES project
(grant agreement \#780751, European Union's Horizon 2020 research and innovation programme)
and MIUR-PRIN 2017 ``Algorithms, Data Structures and Combinatorics for Machine Learning''.
}

\begin{abstract}
The representation of a dynamic ordered set
of $n$ integer keys drawn from a universe
of size $m$
is a fundamental data structuring problem.
Many solutions to this problem achieve optimal time but take
polynomial space, therefore preserving time optimality
in the \emph{compressed} space regime is the
problem we address in this work.
For a polynomial universe $m = n^{\Theta(1)}$,
we give a solution that takes $\ef(n,m) + o(n)$ bits,
where $\ef(n,m) \leq n\lceil \log_2(m/n)\rceil + 2n$
is the cost in bits of the \emph{Elias-Fano} representation
of the set,
and supports random access to the $i$-th smallest element
in $O(\log n/ \log\log n)$ time,
updates and predecessor search in $O(\log\log n)$ time.
These time bounds are optimal.
\end{abstract}



\end{frontmatter}

\section{Introduction}\label{sec:intro}

The \emph{dynamic ordered set problem}
with integer keys is to
represent a set $\seq \subseteq [m] = \{0,\ldots,m-1\}$,
with $|\seq| = n$, such that
the following operations are supported:
$\search(x)$ determines whether $x \in \seq$;
$\ins/\del(x)$ inserts/deletes $x$ in/from $\seq$;
$\pred/\suc(x)$ returns the next smaller/larger
element from $\seq$;
${\minimum}/{\maximum}()$ returns
the smallest/largest element from $\seq$.
This is among the most studied problems in Computer Science
(see the introduction to parts \RNum{3}
and \RNum{5} of the book by~\citet{CLRS}).
Many solutions to this problem are known to require an
optimal amount of time per operation within polynomial space.
For example, under the
comparison-based model that allows
only two keys to be compared in $O(1)$ time,
it is well-known that any self-balancing search tree data
structure, such as AVL or Red-Black, solves the problem
optimally in $O(\log n)$ worst-case time and $O(n)$ words
of space.
(Unless otherwise specified, all logarithms are binary
throughout the article).

However, working with integer keys makes it possible
to beat the $O(\log n)$-time bound with a
RAM model having word size $w = \Theta(\log m)$ bits
~\cite{PT14,Boas75,Willard83,FW93}.
In this scenario, classical solutions include
the \emph{van Emde Boas tree}~\cite{Boas75,Boas77,BKZ77},
\emph{$x/y$-fast trie}~\cite{Willard83} and
the \emph{fusion tree}~\cite{FW93} ---
historically the first data structure
that broke the barrier of $\Omega(\log n)$,
by exhibiting an improved running time of
$O(\log_w n) = O(\log n / \log\log m)$.

In this work, we are interested in preserving the
asymptotic time optimality for the operations
under \emph{compressed space}.
A simple information-theoretic argument~\cite{pagh2001low}
shows that one needs at least
$
\bound(n,m) = \lceil \log {m \choose n} \rceil
= n \log(em/n) - \Theta(n^2/m) - O(\log n)
$
bits to represent $\seq$ ($e = 2.718$ is the base of the
natural logarithm), because there are
${m \choose n}$ possible ways of selecting $n$ integers
out of $m$.
The meaning of this bound is that \emph{any} solution
solving the problem in optimal time \emph{but} taking polynomial space, i.e.,
$O(n^{\Theta(1)} \log m)$ bits,
is actually $\Omega(n \log n)$ bits larger than necessary.

Interestingly, the \emph{Elias-Fano}
representation~\cite{Elias74,Fano71} of the ordered set $\seq$ uses
$\ef(n,m) \leq n\lceil \log(m/n)\rceil + 2n$ bits which is
at most $n \log(m/n) + 3n$ bits.
For $n = o(\sqrt{m})$, we have that
$\bound(n,m) \approx n \log(m/n) + 1.44n$ bits,
showing that Elias-Fano takes $\bound(n,m) + 1.56n$ bits.
We conclude that Elias-Fano is at most $1.56n$ bits away from
the information-theoretic minimum~\cite{grossi2009more}.
We describe Elias-Fano in Section~\ref{sec:ef}.

Given the total order of $\seq$, it is natural to extend
the problem by also
considering the operation {\access} that, given an
index $0 \leq i < n$, returns the $i$-th smallest element
from $\seq$.
(This operation is also known as {\select}.)
It should also be noted that, for any key $x$,
the operation $\search(x)$ can be implemented by
running $\suc(x)$ and checking whether the returned
value is equal to $x$ or not.
Furthermore, it is well-known that {\pred} and {\suc}
have the same complexities and are solved similarly,
thus we only discuss {\pred}.
Lastly, returning the smallest/largest integer
from $\seq$ can be trivially done by storing these
elements explicitly in $O(\log m)$ bits (which is negligible
compared to the space needed to represent $\seq$)
and updating them as needed upon insertions/deletions.
For these reasons, the problem we consider in this
article is formalized as follows.

\begin{problem}{\textbf{Dynamic ordered set with random access} --- }\label{problem}
Given a non-negative integer $m$, represent an ordered set
$\seq \subseteq [m]$ with $|\seq| = n$, such that
the following operations are supported
for any $x$ and $0 \leq i < n$:
\begin{itemize}
\item $\access(i)$ returns the $i$-th smallest
element from $\seq$,
\item $\ins(x)$ sets $\seq = \seq \cup \{x\}$,
\item $\del(x)$ sets $\seq = \seq \setminus \{x\}$,
\item $\pred(x) = \max\{y \in \seq \, | \, y < x\}$.
\end{itemize}
\end{problem}

\paragraph{Our contribution}
In this article we describe a
solution to Problem~\ref{problem}
whose space in bits is expressed
in terms of $\ef(n,m)$ --- the cost of
representing $\seq$ with Elias-Fano ---
and achieves optimal running times.
We consider a unit-cost RAM model with
word size $w = \Theta(\log m)$ bit, allowing
multiplication.
We study the asymptotic behaviour of the data structures,
therefore we also assume, without loss of generality,
that $n$ is larger than a sufficiently big constant~\cite{pagh2001low}.

For the important and practical case
where the integers come from a polynomial
universe of size $m = n^{\Theta(1)}$,
we give a solution that
uses $\ef(n,m) + o(n)$ bits,
thus introducing a \emph{sublinear redundancy}
with respect to $\ef(n,m)$,
and supports:
{\access} in $O(\log n / \log\log n)$ time,
{\ins}, {\del} and {\pred} in $O(\log\log n)$ time.
The time bound for random access under updates
matches a lower bound
given by~\citet*{FS89} for \emph{dynamic selection}.
Dynamic predecessor search, instead, matches a lower bound
given by~\citet*{PT06}.
Our result significantly improves
the space of the best known solution by~\citet*{PT14}
which takes optimal time but polynomial space, i.e.,
$O(n \log m)$ bits.

In Section~\ref{sec:pre} we discuss related work and preliminaries.
The main result is described in Section~\ref{sec:dynamic}.
In Section~\ref{sec:extensible} we develop
a solution that achieves a better update time
under the assumption that
we can only add a key larger than the maximum in the set
(and delete the maximum).

\section{Preliminaries}\label{sec:pre}

In this section we illustrate the context of our work,
whose discussion is articulated in three parts.
We first describe the static Elias-Fano
representation because it is a key ingredient of our solutions.
Then we discuss the results concerning the static predecessor
and dynamic ordered set (and related) problems,
stressing what lower bounds
apply to these problems.
Recall that we use a RAM model with word size $w = \Theta(\log m)$ bits.

\subsection{Static Elias-Fano representation}\label{sec:ef}

\begin{lem}{\textup{Elias-Fano~\cite{Elias74,Fano71}.}}\label{lem:ef}
An ordered set $\seq \subseteq [m]$,
with $|\seq| = n$, can be represented in
$\ef(n,m) + o(n)$ bits
such that {\access} is supported in $O(1)$
and {\pred} in $O(1+\log(m/n))$,
where $\ef(n,m) \leq n\lceil\log(m/n)\rceil + 2n$.
\end{lem}

\paragraph{Space complexity}
Let $\seq[i]$ indicate the $i$-th smallest of $\seq$.
We write each $\seq[i]$ in binary using $\lceil \log m \rceil$ bits.
The binary representation of each integer is then split into two parts:
a \emph{low} part consisting in the right-most
$\ell = \lceil \log(m/n)\rceil$ bits that we call \emph{low bits}
and a \emph{high} part consisting in the remaining $\lceil \log m \rceil - \ell$ bits that we similarly call \emph{high bits}.
Let us call $\ell_i$ and $h_i$ the values of low and high bits
of $\seq[i]$ respectively.
The integers $L = [\ell_0,\ldots,\ell_{n-1}]$ are written explicitly
in $n \lceil \log(m/n)\rceil$ bits and
they represent the encoding of the low parts.
Concerning the high bits, we represent them in \emph{negated unary}
using a bitmap of $n + 2^{\lfloor \log n \rfloor} \leq 2n$ bits as follows. We start from a $0$-valued bitmap $H$ and set the bit
in position $h_i + i$, for $i = 0,\ldots,n-1$.
It is easy to see that the $k$-th unary value of $H$, say $n_k$,
indicates that $n_k$ integers of $\seq$ have high bits equal to
$k$.
For example, if $H$ is $\{$\bit{1110}, \bit{1110}, \bit{10}, \bit{10}, \bit{110}, \bit{0}, \bit{10}, \bit{10}$\}$
(as in Table~\ref{tab:elias_fano}), we have that $H[1] = \bit{1110}$, so we know
that there are 3 integers in $\seq$ having high bits equal to 1.

Summing up the costs of high and low parts, we derive that Elias-Fano
takes $\ef(n,m) \leq n\lceil\log(m/n)\rceil + 2n$ bits.
Although we can opt for an arbitrary split into high and low parts,
ranging from $0$ to $\lceil \log m \rceil$,
it can be shown that $\ell = \lceil \log(m/n) \rceil$
minimizes the overall space of the encoding~\cite{Elias74}.
As explained in Section~\ref{sec:intro}, the space of Elias-Fano
is related to the information-theoretic minimum: it is
at most $1.56n$ bits redundant.



\begin{table}
\centering
\scalebox{0.9}{\begin{tabular}{l ccc|ccc|c|c|cc|c|c|c}
\toprule

$\seq$ & 3 & 4 & 7 & 13 & 14 & 15 & 21 & 25 & 36 & 38 &   & 54 & 62 \\

\midrule

\multirow{3}{*}{\emph{high}}

& \bit{0} & \bit{0} & \bit{0} & \bit{0} & \bit{0} & \bit{0} & \bit{0} & \bit{0} & \bit{1} & \bit{1} & \gray{\bit{1}} & \bit{1} & \bit{1} \\
& \bit{0} & \bit{0} & \bit{0} & \bit{0} & \bit{0} & \bit{0} & \bit{1} & \bit{1} & \bit{0} & \bit{0} & \gray{\bit{0}} & \bit{1} & \bit{1} \\
& \bit{0} & \bit{0} & \bit{0} & \bit{1} & \bit{1} & \bit{1} & \bit{0} & \bit{1} & \bit{0} & \bit{0} & \gray{\bit{1}} & \bit{0} & \bit{1} \\

\cmidrule(lr){1-14}

\multirow{3}{*}{\emph{low}}

& \bit{0} & \bit{1} & \bit{1} & \bit{1} & \bit{1} & \bit{1} & \bit{1} & \bit{0} & \bit{1} & \bit{1} &         & \bit{1} & \bit{1} \\
& \bit{1} & \bit{0} & \bit{1} & \bit{0} & \bit{1} & \bit{1} & \bit{0} & \bit{0} & \bit{0} & \bit{1} &         & \bit{1} & \bit{1} \\
& \bit{1} & \bit{0} & \bit{1} & \bit{1} & \bit{0} & \bit{1} & \bit{1} & \bit{1} & \bit{0} & \bit{0} &         & \bit{0} & \bit{0} \\

\cmidrule(lr){1-14}

$H$ &
\multicolumn{3}{c}{\bit{1110}} & \multicolumn{3}{c}{\bit{1110}} & \multicolumn{1}{c}{\bit{10}} &
\multicolumn{1}{c}{\bit{10}}   & \multicolumn{2}{c}{\bit{110}}  & \multicolumn{1}{c}{\bit{0}}  &
\multicolumn{1}{c}{\bit{10}}   & \multicolumn{1}{c}{\bit{10}} \\

\cmidrule(lr){1-14}

$L$ &
\multicolumn{3}{c}{\bit{011.100.111}} &
\multicolumn{3}{c}{\bit{101.110.111}} &
\multicolumn{1}{c}{\bit{101}} &
\multicolumn{1}{c}{\bit{001}} &
\multicolumn{2}{c}{\bit{100.110}} &
\multicolumn{1}{c}{} &
\multicolumn{1}{c}{\bit{110}} &
\multicolumn{1}{c}{\bit{110}} \\

\bottomrule
\end{tabular}
}

\caption{An example of Elias-Fano encoding.
\label{tab:elias_fano}}

\end{table}

\paragraph{Example}
Table~\ref{tab:elias_fano} shows a graphical example for the sorted
set $\seq$ $=$ $[$3, 4, 7, 13, 14, 15, 21, 25, 36, 38, 54, 62$]$.
The \emph{missing high bits} embody the representation of the fact that using $\lfloor \log_2 n \rfloor$ bits to represent the high part of an integer, we have \emph{at most} $2^{\lfloor \log_2 n \rfloor}$ distinct high parts because not all of them could be present. In Table~\ref{tab:elias_fano}, we have $\lfloor \log_2 12 \rfloor = 3$ and we can form up to $8$ distinct high parts. Notice that, for example, no integer has high part equal to \bit{101} which are, therefore, ``missing'' high bits.

\paragraph{Random access}
A remarkable property of Elias-Fano is that it can be \emph{indexed}
to support {\access} in $O(1)$ worst-case.
The operation is implemented by building an auxiliary data structure
on top of $H$ that answers $\select_{\bit{1}}$ queries.
The answer to a $\select_b(i)$ query over a
bitmap is the
position of the $i$-th bit set to $b$.
This auxiliary data structure is \emph{succinct} in the sense that it is negligibly small in asymptotic terms, compared to $\ef(n,m)$,
requiring only $o(n)$ additional bits~\citep{MN07,Vigna13},
hence bringing the total space of the encoding to $\ef(n,m) + o(n)$ bits.
For a given $i \in [0,n)$, we proceed as follows.
The low bits $\ell_i$ are trivially retrieved as
$L[i\ell, (i + 1)\ell)$.
The retrieval of the high bits is, instead, more
complicated.
Since we write in negated unary how many integers share the same high part,
we have a {\bit{1}} bit for every integer in $\seq$
and a {\bit{0}} for every distinct high part.
Therefore, to retrieve $h_i$, we need to know how many {\bit{0}}s
are present in $H[0, \select_{\bit{1}}(i))$.
This quantity is evaluated on $H$ in $O(1)$ as $\select_{\bit{1}}(i)-i$.
Lastly, re-linking the high and low bits together is as simple as:
$\access(i) = ((\select_{\bit{1}}(i)-i) << \ell) \mid \ell_i$, where $<<$ indicates the left shift operator and $\mid$ is the bitwise OR.

\paragraph{Predecessor search}
The query $\pred(x)$ is supported in $O(1+\log(m/n))$ time as follows.
Let $h_x$ be the high bits of $x$. Then for $h_x > 0$, $i = \select_{\bit{0}}(h_x) - h_x + 1$ indicates that there are $i$ integers in $S$ whose high bits are less than $h_x$. On the other hand, $j = \select_{\bit{0}}(h_x + 1) - h_x$ gives us the position at which the elements having high bits larger than $h_x$ start. The corner case $h_x = 0$ is handled by setting $i = 0$.
These two preliminary operations take $O(1)$.
Now we can conclude the search in the range $\seq[i,j]$, having \emph{skipped} a potentially large range of elements that, otherwise, would have required to be compared with $x$.
The range may contain up
to $u/n$ integers that we search with binary search.
The time bound follows.
In particular, it could be that $\pred(x) < \seq[i]$:
in this case $\seq[i - 1]$ is the element to return if $i > 0$.



\paragraph{Partitioning the representation}
In this article we will use extensively the following
property of Elias-Fano.

\begin{prop}\label{prop:almost_convexity}
Given an ordered set $\seq \subseteq \{0,\ldots,m\}$,
with $|\seq| = n$, let $\ef(\seq[i,j))$
indicate the Elias-Fano representation of $\seq[i,j)$,
for any $0 \leq i < j \leq n$.
Then given an index $k \in [1,n)$,
we have that $\ef(\seq[0,k)) + \ef(\seq^{\prime}[k,n)) \leq \ef(\seq[0,n))$,
where $\seq^{\prime}[l] = \seq[l]-\seq[k-1]+1$, for $l = k,\ldots,n-1$.
\end{prop}

The property tells us that splitting the Elias-Fano encoding of $\seq$
does not increase its space of representation.
This is possible because each segment can be encoded
with a \emph{reduced universe}, by subtracting to each integer
the last value of the preceding segment (the first segment is left
as it is).
Informally, we say that a segment is ``re-mapped'' relatively
to its own universe.
The property can be easily extended to work with an arbitrary number of splits. Let us now prove it.

\begin{proof}
We know that $\ef(\seq[0,n))$ takes $n\phi + n + \lceil m/2^{\phi} \rceil$ bits, where $\phi = \lceil \log(m/n) \rceil$.
Similarly, $\ef(\seq[0,k)) = k\phi_1 + k + \lceil m_1/2^{\phi_1} \rceil$
and $\ef(\seq^{\prime}[k,j)) = (n-k)\phi_2 + (n-k) + \lceil m_2/2^{\phi_2} \rceil$, where $m_1 = \seq[k-1]$ and $m_2 = \seq[n-1] - m_1 + 1$,
are minimized by choosing
$\phi_1 = \lceil \log(m_1/k)\rceil$ and $\phi_2 = \lceil \log(m_2/(n-k))\rceil$.
Any other choice of $\phi_1$ and $\phi_2$ yields a larger cost, therefore:
$\ef(\seq[0,k)) + \ef(\seq[k,n))$ $\leq$ $k\phi + (n-k)\phi + k + (n-k) + \lceil m_1/2^{\phi} \rceil + \lceil m_2/2^{\phi} \rceil$ $\leq$ $n\phi + n + \lceil m/2^{\phi}\rceil$ $=$ $\ef(\seq[0,n))$.
\end{proof}

An important consideration to make is that Property~\ref{prop:almost_convexity}
needs the knowledge of the value $\seq[k-1]$ to work
--- the pivoting element ---
which can be stored in $O(\log m)$ bits.
This means that for small values of $n$ it can happen that the
space reduction does \emph{not} exceed $O(\log m)$ bits.
Since we do not deal with such values of $n$, we
always assume that this is not the case.

\subsection{The static predecessor problem}\label{sec:pred}

\paragraph{Simple solutions}
There are two simple solutions
to the static predecessor problem.
The first uses an array $P[0..m)$ where we store the answers
to all possible queries. In this case $\pred(x) = P[x]$ for any
$x < m$ ($P[0] = -\infty$), thus
the problem is solved in $O(1)$ worst-case time
and $m\lceil\log m\rceil$ bits.
The second solution stores $\seq$ as a sorted array and answers
the queries using binary search,
therefore taking $n\lceil\log m\rceil$ bits
and $O(\log n)$ worst-case time.
Both solutions are unsatisfactory: the first one because of its
space; the second one because of its time.

\paragraph{Lower bounds}
\citet*{Ajtai88} proved the first $\omega(1)$-time lower bound
for polynomial space, i.e., $O(n^{\Theta(1)})$ memory words,
claiming that $\forall \, w$, $\exists \, n$
that gives $\Omega(\sqrt{\log w})$ query time.
~\citet{miltersen1994lower} elaborated on Ajtai's result and also
showed that $\forall \, n$, $\exists \, w$
that gives $\Omega(\sqrt[3]{\log n})$
query time.

For the \emph{dense} case of $m = n (\log n)^{O(1)}$,
~\citet*{pagh2001low} gave a static data structure
taking $\mathcal{B} + o(n)$ bits and
answering membership and predecessor queries in $O(1)$ worst-case time.
(We consider larger universes in this article.)

~\citet*{BF99,BF02} proved two strong bounds for any {cell-probe} data structure.
They proved that $\forall \, w$, $\exists \, n$ that requires
$\Omega(\log w / \log\log w)$ query time and that $\forall \, n$, $\exists \, w$ that requires $\Omega(\sqrt{\log n / \log\log n})$ query time.
They also gave a static data structure achieving
$$O(\min\{\log w / \log\log w, \sqrt{\log n / \log\log n}\})$$ which is, therefore, optimal.

Building on a long line of research,~\citet*{PT06,PT07} finally proved the following optimal (up to constant factors) space/time trade-off.

\begin{thm}{\textup{\citet*{PT06,PT07}.}}\label{thm:pred}
A static data structure representing $n$ integer keys in $z$ bits,
takes time
$$
\Theta\Big(\min\Big\{\frac{\log n}{\log w}, \log \frac{w - \log n}{a}, \frac{\log(w/a)}{\log ( \frac{a}{\log n} \log(w/a) )}, \frac{\log(w/a)}{\log ( \log(w/a) / \log \frac{\log n}{a} )} \Big\}\Big)
$$
to answer a {\pred} query, where $a = \log(z/n)$.
\end{thm}


This lower bound holds for cell-probe, RAM, trans-dichotomous RAM,
external memory and communication game models.
The first branch of the trade-off indicates that, whenever one integer
fits in one memory word, fusion trees~\cite{FW93} are optimal as they
have $O(\log n / \log w)$ query time.
The second branch holds for \emph{polynomial universes},
i.e., when $m = n^\gamma$, for any $\gamma = \Theta(1)$.
In such important case we have that $w = \Theta(\log m) = \Theta(\log n)$,
therefore $y$-fast tries~\cite{Willard83} and
van Emde Boas trees~\cite{Boas75,Boas77,BKZ77}
are optimal with query time $O(\log\log n)$.
The last two bounds of the trade-off, instead, treat the case for \emph{super-polynomial} universes
and are out of scope for this work.

For example, given a space budget of $O(n \polylg n)$ words
we have $a = O(\log\log n)$, thus implying
that $y$-fast tries and van Emde Boas trees are optimal if
$w = O(\polylg n)$ and fusion trees are optimal if
$\log w = \Omega(\sqrt{\log n} \cdot \log\log n)$.





\paragraph{Predecessor queries in succinct space}
We are now interested in determining the optimal running time
of $\pred$ given the Elias-Fano space bound of
$\ef(n,m) + o(n)$ bits from Lemma~\ref{lem:ef},
knowing that the time for dynamic predecessor
with logarithmic update time
can not be better than that of static predecessor
(allowing polynomial space)~\cite{PT14}.

\return
We make the following observation.

\begin{obs}\label{obs:static}
Given any linear-space data structure
supporting {\pred} in $O(t)$
worst-cast time,
an ordered set $\seq \subseteq [m]$
with $|\seq| = n$ can be represented in
$\ef(n,m) + O(n / 2^{ct} \cdot \log m) + o(n)$ bits
such that {\access} is supported in $O(1)$
and {\pred} in $O(t)$ worst-case time,
for any constant $c > 1$.
\end{obs}

We represent $\seq$ with Elias-Fano and (logically)
divide it into $\lceil n / 2^{ct} \rceil$ blocks
of $2^{ct}$ integers each (the last block may contain less integers).
We can solve {\pred} queries in a block in $O(t)$ time
by applying binary search, given that each access is performed
in $O(1)$ time.
The first element of each block (and its position in $\seq$)
is also stored in the linear-space
data structure solving {\pred} in $O(t)$ time.
The space of such data structure is $O(n / 2^{ct} \cdot \log m)$ bits.

\begin{cor}\label{cor:static_pred}
An ordered set $\seq \subseteq [m]$,
with $|\seq| = n$ and $m = n^{\Theta(1)}$, can be represented in
$\ef(n,m) + o(n)$ bits
such that {\access} is supported in $O(1)$
and {\pred} in optimal $O(\min\{1+\log(m/n),\log\log n\})$
worst-case time.
\end{cor}

The linear-space data structure in Observation~\ref{obs:static}
is chosen to be an $y$-fast trie,
whose $t = \log\log n$ query time is
optimal for polynomial universes
(\emph{second} branch of Theorem~\ref{thm:pred}).
The space of the $y$-fast trie is
$O(n / (\log n)^{c-1}) = o(n)$ bits.

Let $m = n^{\gamma}$, for any $\gamma = \Theta(1)$.
The bound $O(\log\log n)$ only depends on $n$, whereas the plain
Elias-Fano bound of $O(1+\log(m/n))$ depends on both $n$ and $m$,
thus varying $\gamma$ only one of the two bounds is optimal.
In fact, we have that $1+\log(m/n) \leq \log\log n$ whenever
$m \leq \frac{n}{2} \log n$, i.e., when $n^{\gamma-1} \leq \frac{1}{2}\log n$. From this last condition we derive that the plain Elias-Fano bound is less than $\log\log n$
when $1 \leq \gamma \leq 1 + \log\log n / \log n$.
When, instead, $\gamma > 1 + \log\log n / \log n$,
the query time $O(\log\log n)$ is optimal and
{exponentially better} than Elias-Fano.
Therefore, $O(\min\{1+\log(m/n),\log\log n\})$
is an accurate characterization of the {\pred}
time bound with $\ef(n,m) + o(n)$ bits.

\return
However for the rest of the discussion, we
assume that $m$ is sufficiently large
so that $\log\log n < 1+\log(m/n)$,
that is $m > \frac{n}{2}\log n$.

\subsection{Dynamic problems}\label{sec:dynamic_problems}

\paragraph{Ordered set problem}
As far as the {\access} operation is \emph{not} supported,
the following results hold.
The van Emde Boas tree~\cite{Boas75,Boas77,BKZ77} is a recursive data structure that maintains $\seq$ in $O(m \log m)$ bits
and $O(\log w)$ worst-case time.
\citet*{Willard83} improved the space bound to $O(n \log m)$ bits
with the $y$-fast trie. (The bound for {\ins}/{\del} is amortized
rather than worst-case).
When polynomial universes are considered,
~\citet*{PT06} proved that
van Emde Boas trees and $y$-fast tries
have an optimal query time
for the dynamic predecessor problem too,
that is $O(\log\log n)$ worst-case.

\citet*{FW93} showed how to solve that dynamic predecessor problem in
$O(\log n / \log\log m)$ time and $O(n)$ space
with the fusion tree.
This data structure is a $B$-tree with branching factor
$B = w^{O(1)}$ that stores in each internal node a \emph{fusion node}
a small data structure able of answering predecessor queries in $O(1)$
for sets up to $B$ integers.

Extending their result to the dynamic predecessor problem,
\citet*{BF99,BF02} proved that any cell-probe data structure
using $(\log m)^{O(1)}$ bits per memory cell and $n^{O(1)}$ worst-case time for insertions, requires $\Omega(\sqrt{\log n / \log\log n})$
worst-case query time.
They also proved that, under a RAM model,
the dynamic predecessor problem can be solved
in
$O(\min\{\log\log n \cdot \log w / \log\log w$, $\sqrt{\log n / \log\log n}\})$,
using linear space.
This bound was matched by~\citet*{AT07} with the so-called
\emph{exponential search tree}.
This data structure has an optimal bound of $O(\sqrt{\log n / \log\log n})$ worst-case time for searching and updating $\seq$, using polynomial space.


\paragraph{Set problems with random access}
The lower bound for the problem changes
by considering the {\access} operation
because this operation is related to the
\emph{partial sums problem} that is,
given an integer array $A[0..n)$,
support $\sm(i)$ returning the sum of the
first $i+1$ integers,
$\upd(i,\Delta)$ which sets $A[i]$ to $A[i] + \Delta$
and $\search(x)$ which returns the index
$i \in [0,n)$ such that $\sm(i) < x \leq \sm(i+1)$.
~\citet*{FS89}
proved a bound of
$\Omega(\log n / \log\log n)$
amortized time for this problem
(see also the extended version of the article by~\citet*{PT14} --- Section 5).
Therefore, this is the lower bound that applies to our
problem as well.
\citet{BCGSVV15} extended the problem as to also support dynamic
changes to the array.

%


\citet*{FS89} also proved that $\Omega(\log n / \log\log n)$
amortized is necessary
for the \emph{list representation problem}, that
is to support {\access}, {\ins} and {\del}.
However, this
problem is slightly different than the one tackled here,
because one can specify the \emph{position} of insertion of a key.
Likewise, the {\del} operation specify the position of the key,
rather than the key itself.
\citet*{RRR01} also addressed the list representation problem
(referred to as the \emph{dynamic array problem})
and provide two solutions.
The first solution is given by the following lemma.

\begin{lem}{\textup{\citet*{RRR01}.}}\label{lem:dynamic_array}
A dynamic array containing $n$ elements
can be implemented to support {\access} in $O(1)$,
{\ins} and {\del} in $O(n^{\epsilon})$ time
using $O(n^{1-\epsilon})$ pointers,
where $\epsilon$
is any fixed positive constant.
\end{lem}

The second solution supports all the three operations
in $O(\log n / \log\log n)$ amortized time.
Both solutions take $o(n)$ bits of redundancy
(besides the space needed to store the array)
and the time bounds are optimal.

Since it takes, $O(B^4)$ time to construct and update a fusion node
with $B$ keys,
~\citet*{PT14} showed that it is possible
to ``dynamize'' the fusion node and obtained the following result.

%
%

\begin{lem}{\textup{\citet*{PT14}.}}\label{lem:optimal_dynamic_sets}
An ordered set $\seq \subseteq [m]$,
with $|\seq| = n$, can be represented in
$O(n \log m)$ bits and supporting
{\ins}, {\del}, {\rank}, {\select} and {\pred}
in $O(\log n / \log\log m)$ per operation.
\end{lem}

The time bound of $O(\log n / \log\log m)$ is optimal, matching
a lower bound by~\citet*{FS89} for dynamic ranking and selection,
and that of predecessor queries for non-polynomial universes
(first branch of the trade-off from Theorem~\ref{thm:pred}).

\section{Succinct Dynamic Ordered Sets with Random Access}\label{sec:dynamic}

In this section we illustrate
our main result for polynomial universes:
a solution to Problem~\ref{problem}
that uses $\ef(n,m) + o(n)$ bits and
supports all operations in optimal time.
From Section~\ref{sec:dynamic_problems}, we recall
that a bound of $\Omega(\log n / \log\log n)$
applies to the {\access}
operation under updates;
{\pred} search needs, instead,
$\Omega(\log\log n)$ time
as explained in Section~\ref{sec:pred}.

\begin{thm}\label{thm:dynamic_opt_ef1}
An ordered set $\seq \subseteq [m]$,
with $|\seq| = n$ and $m = n^{\Theta(1)}$,
can be represented in $\ef(n,m) + o(n)$ bits
such that {\access} is supported in $O(\log n / \log\log n)$,
{\ins}, {\del} and {\pred}
in $O(\log\log n)$ time.
\end{thm}

We first show how to handle small sets of
integers efficiently in Section~\ref{sec:small}.
Then we use this solution to give the
final construction in Section~\ref{sec:large}.

\subsection{Handling small sets}\label{sec:small}

In this section, we give a solution
to Problem~\ref{problem}
working for a small set of integers.

\return
The following lemma is useful.

\begin{lem}{\textup{\citet{JSS12}}}\label{lem:mm}
Given a collection of $k$ blocks, each of size $O(b)$ bits,
we can store it using $O(k \log k + b^2)$ bits
of redundancy to
support {\addr} in $O(1)$ time and {\realloc}
in $O(b/w)$ time.
\end{lem}

We say that the data structure of Lemma~\ref{lem:mm}
has parameters $(k, b)$.
The operation ${\addr}(i)$ returns a pointer
to where the $i$-th block is stored in memory;
the operation ${\realloc}(i,b^\prime)$ changes
the length of the $i$-th block to $b^\prime$ bits.

\return
Now we show the following theorem.

\begin{thm}\label{lem:small_blocks_ds}
Let $\seq \subseteq [m]$ be an ordered set
with $|\seq| = n$ and $m = n^{\Theta(1)}$.
Then a subset $\seq^{\prime}$ of $\seq$,
with $|\seq^{\prime}| = n^{\prime} = \Theta((\log n \cdot \log\log n)^2)$
and $\seq^{\prime} \subseteq [m^{\prime}]$,
can be represented with
$\ef(n^{\prime},m^{\prime}) + O((\log n)^2 \cdot \log\log n) + o(n^{\prime})$
bits and supporting
{\access}, {\ins}, {\del} and {\pred}
in $O(\log\log n)$ time.
\end{thm}

\paragraph{Memory management}
We divide the ordered elements of $\seq^{\prime}$ into blocks
of size $\Theta((\log\log n)^2)$
and represent each block with Elias-Fano.
We have $O((\log n)^2)$ blocks.
Physically, the high and low parts of the Elias-Fano representations
are stored using two different data structures.

The high parts of all blocks are stored using the
data structure of Lemma~\ref{lem:mm},
with parameters $(O((\log n)^2), \Theta((\log\log n)^2))$.
For this choice of parameters, we support
both {\addr} and {\realloc} in $O(1)$ time
and pay a redundancy of
$O((\log n)^2 \cdot \log\log n + (\log\log n)^4)$ $=$
$O((\log n)^2 \cdot \log\log n)$ bits.
This allows to manipulate the high part of a block
in $O(1)$ time upon {\access}, {\ins} and {\del}.

The low parts are stored in a collection of $\Theta((\log\log n)^2)$
dynamic arrays, each being an instance of the data structure
of Lemma~\ref{lem:dynamic_array}.
We maintain an array $A$ of $\Theta((\log\log n)^2)$ pointers to such
data structures, taking $O((\log\log n)^2 \cdot \log n) = O((\log n)^2)$ bits.
Each array stores $O((\log n)^2)$ integers and supports
{\access} in $O(1)$, {\ins} and {\del} in $O(\log\log n)$ as soon
as $\epsilon < 1/6$ in Lemma~\ref{lem:dynamic_array}.
The redundancy to maintain the arrays
is $o(n^{\prime})$ bits.

\paragraph{Indexing}
The blocks are indexed
with a $\tau$-ary tree $\mathcal{T}$,
with $\tau = (\log n)^{\sigma}$ and $0 < \sigma < 1$.
It follows that
the height of the tree is constant and equal to
$h = O(\log_{\tau} (\log n)^2) = O(1/\sigma)$.
The tree operates as a B-tree where
internal nodes store $\Theta(\tau)$ children.
In particular, each node stores
$\Theta(\tau)$ counters, telling how many integers
are present in the leaves that descend from each child.
These counters are kept in prefix-sum fashion to enable
binary search.
Such counters takes $O(\tau \log\log n) = o(\log n)$ bits
which fit in (less than) a machine word.
This allows us to update such counters in $O(1)$ time
upon insertions/deletions.

Each leaf node also stores two offsets per block,
each taking $O(\log\log n)$ bits.
The first offset is the position in $A$ of the pointer
to the dynamic array storing the low parts
of the Elias-Fano representation of the block.
The second offset tells where the low parts of the block
are stored inside the dynamic array.
Thus the overhead per block is $O(\log\log n)$ bits.
As usual, each internal node also stores a pointer
per child, thus maintaining the tree topology
imposes an overhead per block
equal to $O(\log n / \tau) = O((\log n)^{1 - \sigma}) = O(\log\log n)$ bits
as soon as $\sigma \geq 2/3$.
Since the overhead per block is $O(\log\log n)$ bits,
it follows that
the total space of $\mathcal{T}$ is $=$
$O((\log n)^2 \cdot \log\log n)$ bits.

\paragraph{Operations}
To support {\access}, we navigate the tree and spend
$O(\log \tau)$ per level, which is $O(\sigma \log\log n)$, by binary searching
the counters. The proper block is therefore identified in
$O(h \times \sigma \log\log n) = O(\log\log n)$ and the wanted integer is
returned in $O(1)$ time from it knowing the
local offset of the integer inside the block
calculated during the traversal.

To support {\ins}, we need to identify the proper
block where to insert the new integer.
(The {\del} operation is symmetric.)
Again, we use binary search on each level of the tree
but searching among the last values of the indexed blocks.
We can retrieve the last value of a block in $O(1)$,
having the pointer to the block and its size information
from the counters.
This is trivial at the leaves.
In the internal nodes, instead,
if the upper bound of the $i$-th child
is needed for comparison for some $1 \leq i \leq \Theta(\tau)$,
we access the block storing such value by
following the pointer to the \emph{right-most}
block indexed in the sub-tree rooted in the $i$-th child.
Accessing the right-most block takes $O(1)$ time.
Having located the proper block, we insert the new integer in
$O(\log\log n)$ time, as explained before.
Updating the counters in each node of the tree
along the root-to-leaf path takes
$O(1)$ time as they fit in $o(\log n)$ bits.
If a split or merge of a block happens, it is handled as
in a B-tree and solved in a constant number of $O(1)$-time
operations.

During a {\pred} search we identify the proper block
in $O(\log\log n)$ time as explained for {\ins} and return
the predecessor by binary searching the block's values.
The total time of the search is $O(\log\log n)$.

\paragraph{Space complexity}
We now analyze the space taken by the Elias-Fano representations
of the blocks.
Our goal is to show that such space
can be bounded by $\ef(n^{\prime},m^{\prime})$,
that is the space of encoding the set $\seq^{\prime}$ with Elias-Fano.
Since the universe of representation of a block
could be as large as $m^{\prime}$,
storing the lower bounds of the blocks in order to use reduced universes
--- as for Property~\ref{prop:almost_convexity} ---
would require $O((\log n)^3)$ bits of redundancy.
This is excessive because if the data structure is replicated
every $n^{\prime}$ integers to represent a larger
dynamic set $\seq$ with $|\seq| = n$, then these lower bounds would cost
$O(n / (\log\log n)^2 \cdot \log n)$ bits, which is \emph{not}
sub-linear in $n$.
We show that this extra space can be avoided,
observing that the number of bits used to represent
the {low part} of Elias-Fano remains the same for
a sufficiently long sequence of $p$ updates.

From Section~\ref{sec:ef} recall that Elias-Fano
represents each low part with
$\lceil \phi \rceil = \lceil \log(m^{\prime}/n^{\prime}) \rceil$
bits.
Now, suppose that the low parts of the blocks are
encoded using a sub-optimal value $\lceil \mu \rceil$
instead of $\lceil \phi \rceil$.
After we perform $p$ updates,
$\lceil \mu \rceil = \lceil \log(m^{\prime}/(n^{\prime} \pm p)) \rceil$ is set to
$\lceil \phi \rceil$ by rebuilding the blocks.
It is easy to see that $n^{\prime}$ updates are required
to let $\lceil \mu \rceil$ become $\lceil \phi \rceil \pm 1$,
because $\lceil \log(\cdot) \rceil$ changes by $+1$ ($-1$) whenever
its argument doubles (halves).
Therefore we have $\lceil \mu \rceil = \lceil \phi \rceil$
for any $p < n^{\prime}$.
In our case $n^{\prime} = \Theta((\log n \cdot \log\log n)^2)$.
In order to guarantee an amortized cost for
update equal to $O(\log\log n)$, we set $p = O((\log n)^2 \cdot \log\log n)$.
Storing the current value of $\lceil \mu \rceil$ adds
a global redundancy of
$\Theta(\log n)$ bits which is negligible.

\subsection{Final construction}\label{sec:large}

Now we prove the final result -- Theorem~\ref{thm:dynamic_opt_ef1} --
whose key ingredient
is the data structure given in Theorem~\ref{lem:small_blocks_ds}.

\paragraph{Lower level}
We divide the ordered elements of $\seq$ into
blocks of size $\Theta((\log n \cdot \log\log n)^2)$ and represents them
using the tree data structure of Theorem~\ref{lem:small_blocks_ds}.
Therefore, we have a forest $\{\mathcal{T}_i\}$
of $k = \Theta(n / (\log n \cdot \log\log n)^2)$ such data structures.

\paragraph{Upper level}
The first element of each block is
(also) stored in the data structure
of Lemma~\ref{lem:optimal_dynamic_sets}
that is a dynamic fusion tree with out-degree $\Theta(\log n)$,
and in an $y$-fast trie.
Let call these data structures $\mathcal{F}$ and $\mathcal{Y}$
respectively.
The $i$-th leaf of both $\mathcal{F}$ and $\mathcal{Y}$
holds a pointer to the
data structure $\mathcal{T}_i$.

\paragraph{Space and time complexity}
The lower level costs
$O(k \cdot (\log n)^2 \cdot \log\log n) + o(n)$
$=$ $O(n / \log\log n) + o(n) = o(n)$ bits.
The total cost of the upper level is
$O(k \cdot \log n) = O(n / (\log n \cdot (\log\log n)^2)) = o(n)$ bits.
Since each block is re-mapped relatively to its universe,
Property~\ref{prop:almost_convexity} guarantees that
the space of representation is at most $\ef(n,m)$ bits.
The space bound claimed in Theorem~\ref{thm:dynamic_opt_ef1} follows.

A total running time of $O(\log n/ \log\log n)$
for {\access} follows
because the $\mathcal{F}$ data structure
operates in this time.
For {\ins}, {\del} and {\pred}, we use the $\mathcal{Y}$
data structure, thus attaining to $O(\log\log n)$ time.
(The bound for {\ins} and {\del} is amortized rather than worst-case).

\section{Append-only}\label{sec:extensible}

In this section we extend the
result given in Corollary~\ref{cor:static_pred}
to the case where
the integers are inserted in sorted order using
an {\append} operation.
In this case, we obtain an append-only representation.


\begin{thm}\label{thm:ext_ef}
An ordered set $\seq \subseteq [m]$,
with $|\seq| = n$ and $m = n^{\Theta(1)}$,
can be represented in $\ef(n,m) + o(n)$ bits such that
{\append} and {\access} are supported in $O(1)$ time,
{\pred} in $O(\log\log n)$ time.
\end{thm}

\paragraph{Data structure and space analysis}
We maintain an array $A[0..k)$ of size $k = O((\log n)^c)$ where
integers are appended uncompressed, for any $c > 1$.
The array is periodically encoded with Elias-Fano in $\Theta(k)$ time
and overwritten.
Each compressed representation of the buffer is appended to
another array of blocks encoded with Elias-Fano.
More precisely, when $A$ is full we encode with Elias-Fano its corresponding \emph{differential} buffer, i.e., the buffer whose values are $A[i] - A[0]$, for $0 \leq i < k$.
Each time the array is compressed, we append in another array
$A^{\prime}$ the pair
(\var{base}, \var{low}) $=$ $(A[0], \lceil\log(A[k-1]/k)\rceil)$, i.e., the buffer lower bound value (\var{base}) and the number of bits (\var{low}) needed to encode the average gap of the Elias-Fano representation of the block.

As discussed for Corollary~\ref{cor:static_pred},
we store the buffer lower bounds an $y$-fast trie.
More precisely, it stores a buffer lower bound
and the index of the Elias-Fano-encoded block to which
the lower bound belongs to.
The space of this data structure is $o(n)$ bits.
Besides the space of the $y$-fast trie, which is $o(n)$ bits,
and that of the Elias-Fano-encoded blocks,
the redundancy of the data structure is due to
(1) $O((k+1)\log n)$ bits for the array $A$ and its (current) size;
(2) $O(n/k \cdot \log n)$ bits for pointers to the Elias-Fano-encoded
blocks;
(3) $O(n/k \cdot \log n)$ bits for the array $A^\prime$;
and it sums up to $o(n)$ bits.

Lastly,
Property~\ref{prop:almost_convexity} guarantees that the space taken by the blocks encoded with Elias-Fano can be safely upper bounded by $\ef(n,m)$
so that the overall space of the data structure is at most
$\ef(n,m) + o(n)$ bits.

\paragraph{Operations}
The operations are supported as follows.
Since we compress the array $A$ each time it fills up
(by taking $\Theta(k)$ time), {\append} is performed in $O(1)$ amortized time.
Appending new integers in the buffer accumulates a credit of $\Theta(k)$
that (largely) pays the cost $O(\log\log n)$
of appending a value to the $y$-fast trie.
To {\access} the $i$-th integer, we retrieve the element $x$ in position
$i - p \times k$ from the compressed block of index $p = \lfloor i/k \rfloor$.
This is done in $O(1)$ worst-case time, since we know how many low bits are required to perform {\access} by reading $C[p].\var{low}$.
We finally return the integer $x + C[p].\var{base}$.
To solve $\pred(x)$, we first resolve a partial $\pred(x)$ query
in the $y$-fast trie to identify
the index $k$ of the compressed block
where the predecessor is located.
This takes $O(\log\log n)$ worst-case time.
We return $C[p].\var{base} + \pred(x - C[p].\var{base})$
by binary searching the block of index $p$ in $O(\log\log n)$
worst-case time.

\section{Conclusions}

In this paper we have shown that Elias-Fano
can be used to obtain
a succinct dynamic data structure with
optimal update and query time,
solving the \emph{dynamic ordered set with random access}
problem.
Our main result holds for
polynomial universes and
is a data structure
using the same asymptotic space of Elias-Fano
--- $\ef(n,m) + o(n)$ bits, where
$\ef(n,m) \leq n\lceil \log(m/n) \rceil + 2n$ ---
and supporting {\access}
in $O(\log n / \log\log n)$ time,
{\ins}, {\del} and {\pred}
in $O(\log\log n)$ time.
All time bounds are optimal.
Note that the space of the solution can be rewritten
in terms of information-theoretic minimum
$\bound(n,m) = \lceil \log {m \choose n} \rceil$
since $\ef(n,m) = \bound(n,m) + 1.56n$ bits.

\return
An interesting open problem is:
\emph{Can the space be improved
to $\bound(n,m) + o(n)$ bits
and preserving the operational bounds?}

\return
Another question is: \emph{Can the result be extended to
non-polynomial universes?}

\return
In this case, the lower bound for dynamic predecessor search
is $O(\log_w n) = O(\log n / \log\log m)$ that corresponds to
the \emph{first} branch
of the time/space trade-off in Theorem~\ref{thm:pred},
as well as the one for {\access}, {\ins} and {\del}~\cite{PT14}.
It seems that a different solution than the
one described here has to be found since
the data structure of Theorem~\ref{thm:dynamic_opt_ef1}
allows us to support all operations in time $O(\log\log m)$
when non-polynomial universes are considered.
Therefore, we give the following
corollary that matches the asymptotic time bounds of $y$-fast tries and
van Emde Boas trees (albeit sub-optimal)
but in almost optimally compressed space.

\begin{cor}
An ordered set $\seq \subseteq \{0,\ldots,m\}$,
with $|\seq| = n$, can be represented in
$\ef(n,m) + o(n)$ bits
such that
{\access}, {\ins}, {\del} and {\pred} are all
supported in $O(\log\log m)$ time.
\end{cor}


\bibliographystyle{plainnat}
\bibliography{bibliography}

\end{document}